\documentclass[conference]{IEEEtran}
\IEEEoverridecommandlockouts
\usepackage[T1]{fontenc}
\usepackage{inconsolata}
\usepackage{amsmath,amssymb}
\usepackage{newtxmath}
\usepackage{graphicx}
\usepackage{xcolor}
\usepackage{tikz}
\usepackage{eso-pic}
\usetikzlibrary{arrows.meta}
\usepackage[nobreak]{cite}
\usepackage{hyperref}
\usepackage{fontawesome5}
\usepackage{xurl}
\urldef\UrlDruid\url{https://github.com/apache/druid/compare/dd20950^...a39eb65}
\urldef\UrlSurvey\url{https://survey.stackoverflow.co/2025/technology#1-dev-id-es}

\usepackage{xspace}
\usepackage{multirow}
\usepackage{cuted}

\makeatletter
  \let\MYcaption\@makecaption
\makeatother
\usepackage{subcaption}
\makeatletter
  \let\@makecaption\MYcaption
\makeatother
\newcommand{\Heading}[1]{\textbf{#1.}}
\newcommand{\RQ}[1]{\textit{RQ}\textsubscript{#1}}
\usepackage{framed}
\newcommand{\Conclusion}[1]{\begin{framed}\noindent #1\end{framed}}

\def\MyToolName{D-Diff\xspace}

\def\NumParticipants{8\xspace}
\def\NumProblemsTotal{8\xspace}
\def\NumProblemsPerGroup{4\xspace}
\def\NumProblemsIntraLine{4\xspace}
\def\NumCommitSequences{1{,}410\xspace}
\def\NumCommitPairs{536\xspace}
\def\NumCommitPairsSingleFile{30\xspace}
\def\BoundaryAlterationRate{20\xspace}
\def\MedianTimeBaseline{530\xspace}
\def\MedianTimeDdiff{230\xspace}
\def\TimeReductionSec{300\xspace}
\def\TimeReductionPct{57}
\def\MeanScreenSwitches{18.6\xspace}
\def\MinProblemMedianSwitches{13\xspace}
\def\MaxProblemMedianSwitches{25\xspace}

\def\NumParticipantsRecorded{seven\xspace}
\def\PValueAccuracy{0.5469\xspace}
\def\PValueEfficiency{0.0078\xspace}
\def\CliffsDelta{0.8438\xspace}
\def\CliffsDeltaInterpret{large\xspace}
\def\PValueAccuracyPerTask{0.9165\xspace}
\def\PValueEfficiencyPerTask{0.0156\xspace}

\usepackage[noabbrev]{cleveref}
\crefname{figure}{Fig.}{Figs.}
\Crefname{figure}{Figure}{Figures}
\crefname{table}{Table}{Tables}
\crefname{section}{Section}{Sections}

\begin{document}

\title{{\spaceskip=0.74\fontdimen2%
\MyToolName: An Interactive Environment for Adjusting Commit Boundaries Based on an Editable 3-way Diff}}

\author{%
  \IEEEauthorblockN{Daiki Muto, Takayoshi Ueno, Shinpei Hayashi}
  \IEEEauthorblockA{%
    \textit{School of Computing, Institute of Science Tokyo}, Tokyo 152--8550, Japan\\
    \{muto, ueno, hayashi\}@se.comp.isct.ac.jp
  }
}
\maketitle
\pagestyle{plain}
\thispagestyle{plain}

\AddToShipoutPictureBG*{%
  \AtPageLowerLeft{\raisebox{18.6mm}{%
    \hspace{\dimexpr\oddsidemargin+1in\relax}%
    \fbox{\parbox[b]{\dimexpr\textwidth-2\fboxsep-2\fboxrule\relax}{\tiny
      \copyright~2026 IEEE. Personal use of this material is permitted.
      Permission from IEEE must be obtained for all other uses, in any current or future media,
      including reprinting/republishing this material for advertising or promotional purposes,
      creating new collective works, for resale or redistribution to servers or lists,
      or reuse of any copyrighted component of this work in other works.}}}}}

\begin{strip}\centering%
  \captionsetup{type=figure}%
  \begin{subfigure}[b]{0.49\linewidth}\centering%
    \includegraphics[width=\linewidth]{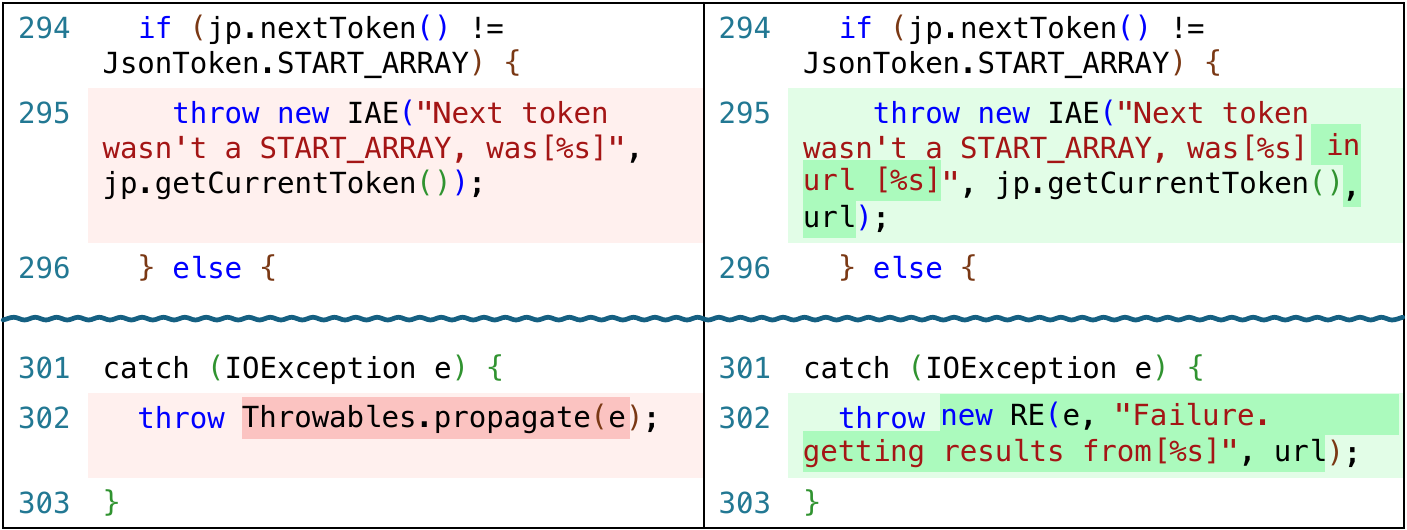}%
    \caption{The first commit shown in a typical diff editor.}
    \label{fig:commit1}
  \end{subfigure}\hfill
  \begin{subfigure}[b]{0.49\linewidth}\centering%
    \includegraphics[width=\linewidth]{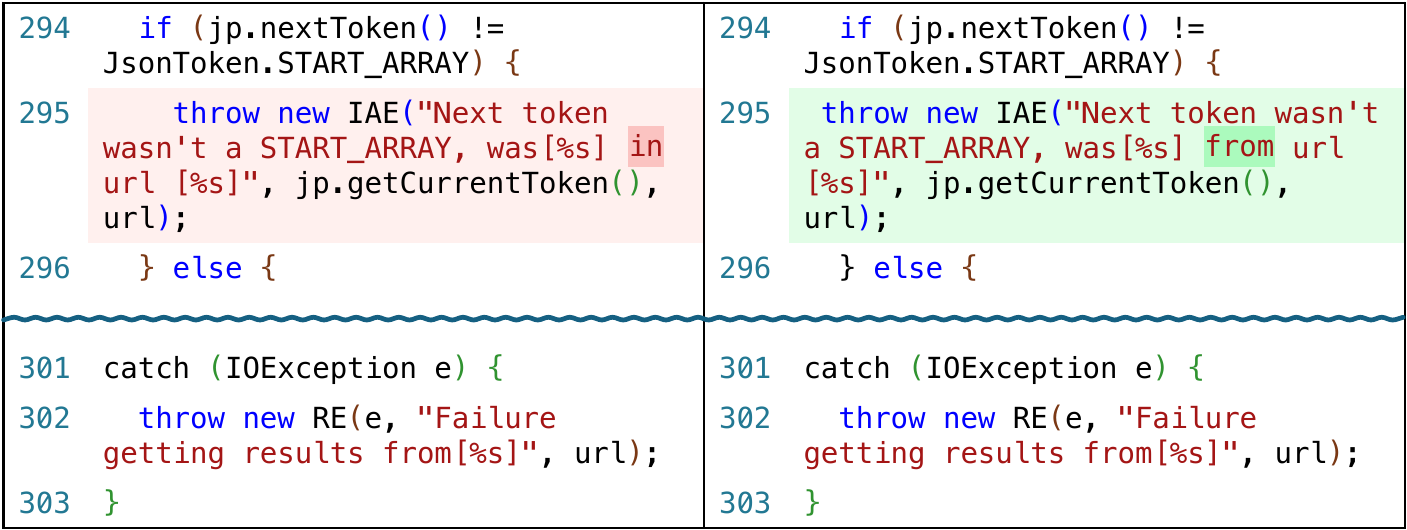}%
    \caption{The second commit shown in a typical diff editor.}
    \label{fig:commit2}
  \end{subfigure}

  \vspace{\baselineskip}
  \begin{subfigure}[b]{\linewidth}\centering
    \includegraphics[width=.9\linewidth,keepaspectratio]{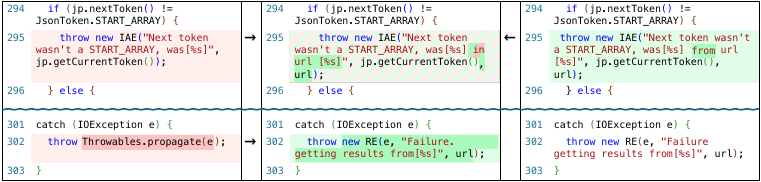}%
    \caption{3-way diff display of \MyToolName.}
    \label{fig:method-overview}
  \end{subfigure}
  \caption{Example of two consecutive commits with a partially altered boundary: (a, b) their diffs displayed separately and (c) integrated by \MyToolName.}\label{fig:approach-overview}
\end{strip}

\begin{abstract}
In version control, it is recommended that each commit include only changes related to one task.
To follow this recommendation, developers may need to adjust commit boundaries, that is, to compare and modify the diffs between two consecutive commits.
Existing tools either display only a single diff at a time, forcing developers to rely on their memory when comparing diffs, or display three files simultaneously without showing the two consecutive diffs, forcing developers to infer them; both increase their cognitive load and hamper the adjustment.
As a first step toward supporting this process, we propose \MyToolName, an interactive diff adjustment environment for two consecutive commits that each involve the same single file.
Based on a 3-way diff display, \MyToolName integrates the two diffs into a single compact view, which also provides a way to modify them.
A within-subjects user study with \NumParticipants participants showed that \MyToolName significantly outperforms a baseline tool in terms of efficiency, reducing the median adjustment time from \MedianTimeBaseline seconds to \MedianTimeDdiff seconds (approximately \TimeReductionPct\%).
\MyToolName also received a higher percentage of positive responses across all usability questionnaire items, while no statistically significant difference was found in accuracy.

\faIcon{video} Video demonstration: \url{https://youtu.be/ra47jbrP5ZA}
\end{abstract}


\section{Introduction}\label{s:introduction}

In version control, it is recommended that each commit include only changes related to one task; such a commit is called a Task Level Commit (TLC) \cite{Berczuk2002}.
TLCs offer advantages such as facilitating the understanding of changes during code reviews \cite{Tao2012} and reducing noise in repository mining \cite{Herzig2013}.
However, developers often create commits that mix multiple intentions \cite{Herzig2013, Tao2015}.
Such commits, called Composite Commits (CC), hinder the understanding, reuse, and reverting of changes \cite{Hayashi2010}.

To follow this recommendation, developers may need to adjust commit boundaries.
In this paper, \textit{adjusting commit boundaries} refers to the process of comparing and modifying the diffs between two consecutive commits.
This process is part of restructuring the entire commit history.
A typical scenario is that a developer adjusts the boundaries of local commits before pushing them to a shared repository, since rewriting already shared history is discouraged.
In this process, developers can leverage approaches that automatically split CCs into TLCs \cite{Fan2024, Hou2025, Zhu2026}, increasingly driven by large language models, to obtain initial commit boundaries.
However, what counts as a single task varies across organizations and projects, so an automatic split rarely matches the developer's intent out of the box; a human must still inspect it and shift changes across commit boundaries to finalize it.
Automatic splitting and manual adjustment are complementary, but the interactive support for this last-mile correction remains limited.

While several tools \cite{Sothornprapakorn2018, Yamashita2020, Shen2021} and standard methods provided by version control systems (VCSs) can be used for adjusting commit boundaries, they are not well-suited for this process.
Most of them display only a single diff at a time, as illustrated in \cref{fig:commit1,fig:commit2}, requiring developers to memorize the not-displayed diff when comparing diffs.
The others display three files simultaneously but show only where the files differ, requiring developers to infer the two consecutive diffs from the displayed differences.

In this paper, to address this challenge, we propose \MyToolName, an interactive diff adjustment environment.
Instead of manipulating commits indirectly through rebase commands and hunk-by-hunk staging, \MyToolName lets developers directly edit the state that lies between the two commits: it shows the two diffs around this shared boundary version using a 3-way diff display, as shown in \cref{fig:method-overview}, and the commits are updated as the developer moves changes across the boundary or edits it.
This simultaneous view reduces the cognitive load of comparing diffs.
As a first step toward supporting this process, \MyToolName focuses on cases where the two commits each involve the same single file.

The primary contributions of this paper are as follows:
\begin{itemize}
  \item We characterize adjusting commit boundaries as the comparison and modification of the diffs between two consecutive commits, and show that existing tools are ill-suited to it because they either display only a single diff at a time, forcing developers to rely on their memory when comparing diffs, or display three files simultaneously without showing the two consecutive diffs, forcing developers to infer them.
  \item We propose \MyToolName, an interactive diff adjustment environment for two consecutive commits that each involve the same single file.
        \MyToolName integrates, in one screen, (i) the simultaneous comparison of two diffs by adapting a 3-way diff display, with a coloring scheme that distinguishes the changes of each commit from those shared by both, and (ii) the modification of diffs in multiple ways: line-level movement, hunk-level movement, and direct editing.
        The alignment reuses established differencing techniques rather than a new algorithm; the contribution is the interaction design that turns editing the shared boundary state into a commit-boundary operation.
  \item We provide empirical evidence, through a within-subjects user study with \NumParticipants participants, that \MyToolName significantly improves the efficiency of adjusting commit boundaries over a baseline tool (median time reduced from \MedianTimeBaseline to \MedianTimeDdiff seconds, about \TimeReductionPct\%), and is rated higher in usability, while no statistically significant difference in accuracy is observed.
\end{itemize}

The structure of this paper is as follows.
\Cref{s:motivation} clarifies the requirements for the proposed approach.
\Cref{s:related-work} reviews related work.
\Cref{s:technique} presents the design and implementation of \MyToolName.
\Cref{s:evaluation} evaluates it through a user study.
\Cref{s:conclusion} concludes.

\section{Requirements for the Proposed Approach}\label{s:motivation}

Adjusting commit boundaries consists of comparing the contents of diffs included in two commits and modifying the diffs so that changes are included in the correct commit.
Developers need to examine both commits simultaneously to determine which commit each change should belong to, and to check how the boundary changes as they modify the diffs.
Given three versions of a file, $s$, $s'$, and $s''$, the first commit is represented by the diff between $s$ and $s'$, and the second commit is represented by the diff between $s'$ and $s''$.
In this view, the intermediate version $s'$ is the boundary shared by the two commits, and modifying the boundary can be modeled as editing this intermediate version.

\def\X{$\checkmark$}
\begin{table}[tb]
  \centering
  \caption{Comparison of Existing Tools Usable\\for Adjusting Commit Boundaries}\label{t:existing-tools}
  {\tabcolsep=2pt\begin{tabular}{llcc}
    \hline
    & & \multicolumn{2}{c}{\!\!\!\!Modification Method} \\
    Tool & Comparison Method & Editing & Moving \\\hline
    git command & Single diff + memory & \X & \\
    GUI clients and IDEs & Single diff + memory & \X & \X \\
    \begin{tabular}[t]{@{}l@{}}Jujutsu \cite{jujutsu} +\\\ \ Meld \cite{meld} / Diffedit3 \cite{diffedit3}\end{tabular} & Single diff + memory & \X & \X \\
    ChTree \cite{Sothornprapakorn2018} & Single diff + memory & & \X \\
    ChangeBeadsThreader \cite{Yamashita2020} & Single diff + memory & & \X \\
    SmartCommit \cite{Shen2021} & Single diff + memory & & \X \\
    diff3 \cite{diff3} & Three files simultaneously & & \\
    vimdiff (Vim \cite{vimeditor}) & Three files simultaneously & \X & \X \\\hline
  \end{tabular}}
\end{table}

\begin{figure}[tb]
  \centering
  \begin{tikzpicture}[
    font=\scriptsize,
    win/.style={rounded corners=2pt, draw=black!70, fill=white, line width=0.5pt},
    winghost/.style={rounded corners=2pt, draw=black!50, dashed, fill=black!6, line width=0.5pt},
    titlebar/.style={fill=black!12, draw=none},
    ctx/.style={black!30, line width=1.2pt, line cap=round},
    del/.style={red!55, line width=1.2pt, line cap=round},
    add/.style={green!55!black!50, line width=1.2pt, line cap=round},
    flow/.style={-{Stealth[length=2mm]}, line width=0.6pt, black!70},
    swap/.style={{Stealth[length=2mm]}-{Stealth[length=2mm]}, line width=0.7pt, black!70},
    note/.style={black, align=center},
  ]

  \draw[winghost] (4.6,1.0) rectangle (8.60,3.1);
  \fill[black!12] (4.6,2.65) rectangle (8.60,3.1);
  \draw[winghost, fill=none] (4.6,1.0) rectangle (8.60,3.1);
  \node[anchor=west, black] at (4.7,2.87) {Diff of the first commit ($s {\leftrightarrow} s'$)};

  \draw[win] (4.3,0.3) rectangle (8.30,2.4);
  \fill[titlebar] (4.3,1.95) rectangle (8.30,2.4);
  \draw[win, fill=none] (4.3,0.3) rectangle (8.30,2.4);
  \node[anchor=west] at (4.4,2.17) {Diff of the second commit ($s' {\leftrightarrow} s''$)};
  \draw[black!25, line width=0.4pt] (6.4,0.3) -- (6.4,1.95);
  \draw[ctx] (4.6,1.72) -- (6.1,1.72);
  \draw[ctx] (4.6,1.43) -- (5.7,1.43);
  \draw[del] (4.6,1.14) -- (5.9,1.14);
  \draw[ctx] (4.6,0.85) -- (6.0,0.85);
  \draw[ctx] (4.6,0.56) -- (5.5,0.56);
  \draw[ctx] (6.55,1.72) -- (7.95,1.72);
  \draw[ctx] (6.55,1.43) -- (7.65,1.43);
  \draw[add] (6.55,1.14) -- (7.98,1.14);
  \draw[ctx] (6.55,0.85) -- (7.85,0.85);
  \draw[ctx] (6.55,0.56) -- (7.45,0.56);

  \draw[swap] (8.05,2.45) to[bend right=40] (8.35,3.15);
  \node[note, anchor=east] at (8.55,3.45) {switches repeatedly};
  \node[note] at (6.3,-0.05) {\itshape only one diff is displayed at a time};

  \draw[win] (0.3,3.0) rectangle (3.4,4.3);
  \fill[titlebar] (0.3,3.85) rectangle (3.4,4.3);
  \draw[win, fill=none] (0.3,3.0) rectangle (3.4,4.3);
  \node[anchor=west] at (0.4,4.07) {Interactive rebase};
  \node[anchor=west, black, font=\scriptsize\ttfamily] at (0.5,3.62) {pick~~e9511d2 ...};
  \node[anchor=west, black, font=\scriptsize\ttfamily] at (0.5,3.32) {edit~~a39eb65 ...};

  \fill[black!55] (1.2,0.95) circle (0.24);
  \fill[black!55] (0.8,0.25) to[out=85,in=180] (1.2,0.62) to[out=0,in=95] (1.6,0.25) -- cycle;
  \node[note] at (1.2,-0.05) {developer};

  \fill[black!40] (1.7,1.45) circle (0.035);
  \fill[black!40] (2.0,1.65) circle (0.05);
  \fill[black!40] (2.3,1.85) circle (0.065);
  \draw[black!50, fill=white, line width=0.5pt] (2.9,2.35) ellipse (1.3 and 0.62);
  \draw[ctx] (2.2,2.6) -- (3.2,2.6);
  \draw[del] (2.2,2.42) -- (3.0,2.42);
  \draw[add] (2.2,2.24) -- (3.45,2.24);
  \node[note, font=\tiny] at (2.9,2.0) {memorized first diff};

  \draw[flow] (1.55,0.95) -- node[below=1pt, note] {compares and\\modifies} (4.2,1.2);
  \draw[flow] (1.1,1.3) to[bend left=10] node[left=2pt, note, align=right] {performs\\the rebase} (1.3,2.95);
\end{tikzpicture}%
  \caption{Workflow of adjusting commit boundaries with VS Code and GitLens.}
  \label{fig:baseline-workflow}
\end{figure}
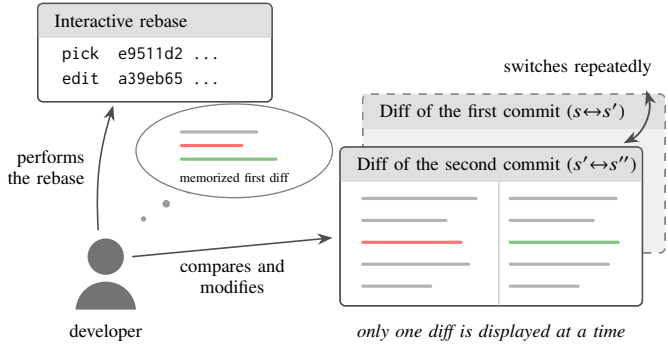

As a concrete example, \cref{fig:baseline-workflow} shows the workflow of a developer adjusting the boundary between two consecutive commits with Visual Studio Code (VS Code)\cite{vscode} and the GitLens\cite{gitlens} extension, the setup we adopt as the baseline in our evaluation (\cref{s:evaluation}).
Modifying the commit to which a change belongs requires an interactive rebase.
More fundamentally, to decide which changes belong to each commit, the developer must compare the diffs of the two commits, yet only one of them can be displayed at a time.
The developer therefore opens the first commit's diff, memorizes the relevant lines, switches to the second commit's diff, and switches back, relying on memory because the two diffs are never visible together.
This comparison recurs throughout the task.

As \cref{t:existing-tools} shows, several tools can be used to adjust commit boundaries, but none of them is well-suited to it.
These tools other than diff3 \cite{diff3} and vimdiff modify diffs by editing changes, moving them, or both; notably, Jujutsu \cite{jujutsu} combined with the Meld \cite{meld} or diffedit3 \cite{diffedit3} diff editors provides an editable pane placed between the two versions of a diff.
However, they display only the diff of one commit at a time.
Developers using them therefore cannot see the two diffs simultaneously, nor observe how the commit boundary changes as they modify the diffs, and must instead rely on their memory.
Conversely, diff3 and vimdiff display three files simultaneously.
However, both of them only show where the files differ from each other: diff3 outputs the differing ranges of lines, and vimdiff highlights them.
Developers must therefore still infer the two consecutive diffs against the middle version.
We argue that these limitations impair work efficiency and may lead to incorrect judgments.

To overcome these limitations, the proposed approach must satisfy two requirements.
\begin{itemize}
  \item \textbf{Requirement~1:} The approach must allow developers to compare the diffs without relying on their memory.
  \item \textbf{Requirement~2:} The approach must provide a method for modifying diffs.
\end{itemize}

\section{Related Work}\label{s:related-work}

\Heading{Tools for Modifying Commit History}
Git and the tools built around it provide general-purpose means of modifying commit history.
Using interactive rebase with \texttt{git rebase -i}, developers can reorder, combine, and split commits.
Furthermore, \texttt{git add -p} allows developers to interactively stage changes on a hunk-by-hunk basis.
GUI clients such as
  GitKraken\cite{gitkraken},
  GitHub Desktop\cite{githubdesktop}, and
  Sourcetree\cite{sourcetree}
are widely used to improve the usability of these git commands.
Some tools further support moving changes across existing commit boundaries.
git absorb \cite{gitabsorb} automatically distributes staged changes as fixup commits to the earlier commits that last modified the corresponding lines.
Jujutsu \cite{jujutsu}, a Git-compatible VCS, provides commands such as \texttt{jj squash} that move changes from one commit into another.
However, all of these general-purpose tools display only one commit at a time, so developers cannot check how the boundary between two consecutive commits changes as they operate.

Apart from these general-purpose tools, several TLC creation support tools have been proposed.
Sothornprapakorn et al.\ proposed a method to visualize changes in a tree structure and implemented ChTree \cite{Sothornprapakorn2018}.
Yamashita et al.\ proposed a method to untangle changes by splitting and merging automatically generated change clusters and implemented ChangeBeadsThreader \cite{Yamashita2020}.
Shen et al.\ proposed SmartCommit \cite{Shen2021}, which performs graph-based clustering and allows developers to edit the results through two types of operations: adjusting thresholds and moving changes between clusters.
These tools are designed to split a single set of changes into groups for creating TLCs.
They display only the single diff being decomposed, and their modification is limited to regrouping changes.

\Heading{Change Untangling}
Many approaches have been proposed to automatically untangle a CC, i.e., to split it into groups of changes, each of which is related to a single task.
Some approaches group changes based on static analysis or heuristic rules \cite{Barnett2015, Tao2015, Muylaert2018, Wang2019}.
Dias et al.\ applied machine learning to fine-grained change histories recorded in the IDE \cite{Dias2015}.
More recent approaches construct graph representations of changes and apply clustering \cite{Partachi2020, Shen2021, Li2022, Chen2022, Fan2024}, or leverage large language models \cite{Hou2025, Zhu2026}.
Guo and Song proposed ChgCutter, which interactively decomposes a CC so that a subset of the changes can be reviewed and tested in isolation \cite{Guo2017}.
While these approaches untangle a CC after it has been created, Kirinuki et al.\ proposed a method that warns developers of tangled changes at commit time to prevent CCs \cite{Kirinuki2014}.
These approaches split a single CC into TLCs, whereas \MyToolName supports adjusting the boundary between two consecutive commits that have already been separated.
They are therefore complementary: automatically untangled results that do not match a project's TLC standards can be adjusted with \MyToolName.

\Heading{Diff Visualization}
A line of research enriches the conventional text-based diff to help developers comprehend changes.
Some approaches augment the diff with the semantics of the change.
RefactorInsight \cite{Kurbatova2021} and RAID \cite{Brito2021} annotate the diff with detected refactorings, in the IDE and on GitHub, respectively.
ChangePrism \cite{Chen2025} overlays refactorings and micro-changes on the diff.
ChangeViz \cite{Gasparini2021} adds method call information to the GitHub pull request diff.
Others refine the diff itself.
Spike \cite{Escobar2022} highlights fine-grained intra-line changes in the editor.
DiffViz \cite{Frick2018} visualizes edit scripts independently of the underlying differencing algorithm, supporting granularities up to the abstract syntax tree level.
These approaches enrich a single diff, whereas \MyToolName displays two consecutive diffs simultaneously to support adjusting commit boundaries.

\Heading{3-way Diff Display}
3-way diff display is a method that presents three versions of source code simultaneously.

Some general-purpose tools can display three files simultaneously.
The diff3 command \cite{diff3} compares three files and outputs the ranges of lines that differ among them.
However, it provides no means of editing and thus cannot be used to adjust commit boundaries.
The vimdiff command, a diff display mode of the Vim \cite{vimeditor} editor, accepts three files and displays them in a Side-by-Side Diff format, allowing the developer to edit the displayed files and to copy a differing hunk from one file to another.
However, its highlighting only shows where each file differs from the other two, not the two consecutive diffs against the middle version.
Developers must therefore infer which commit each difference belongs to.
Moreover, it lacks support for preserving the two outer versions and for integrating the edits into the commit history.

The most widely used form of 3-way diff display is the merge editor.
For example, GUI clients and IDEs such as VS Code provide such editors.
They typically place the two conflicting source codes side by side in the upper panes and display the merge result in the lower pane.
These editors, however, assume two changes diverging from a common base, not two consecutive diffs in a linear history.

More recently, another type of 3-way diff display has emerged for modifying commit history: the two versions of a diff are shown in the left and right panes, and an editable pane is placed in the middle.
Jujutsu experimentally supports this style of editing with two external diff editors.
One is the Meld \cite{meld} merge tool, whose merge view is repurposed for diff editing.
The other is diffedit3 \cite{diffedit3}, a browser-based editor built for this purpose.
These editors are invoked from commands such as \texttt{jj squash -i}, which moves selected changes between two commits.
The developer selects the changes to be moved by editing the middle pane directly or by moving hunks into it from the side panes.
These editors share with \MyToolName the idea of directly editing an intermediate version.
However, the three panes present only the diff of the single commit from which the changes are moved.
The two consecutive diffs that result from the operation are never displayed together with their shared boundary version.
The developer must inspect the resulting diffs one by one after the operation, e.g., with \texttt{jj diff}.

\section{\MyToolName: A Diff Adjustment Environment}\label{s:technique}

In this section, we propose \MyToolName, an interactive diff adjustment environment that satisfies the requirements mentioned above.
In this paper, as a first step toward solving the problem of adjusting commit boundaries, we limit the scope to cases where the two commits each involve the same single file.
This single-file, two-commit case is the building block of the general setting: as long as adjusting the boundary does not require comparing changes across different files, a multi-file commit decomposes into independent per-file adjustments, and a longer commit sequence into consecutive pairs.

\subsection{Overview}\label{s:approach-overview}

\Cref{fig:method-overview}, introduced in \cref{s:introduction}, shows an example of applying \MyToolName to two TLCs\footnote{\UrlDruid} with partially altered commit boundaries.
As shown in the figure, \MyToolName simultaneously visualizes two consecutive diffs of the same file.
This enables developers to compare the diffs without relying on their memory, satisfying Requirement~1.
In addition, \MyToolName provides two operations to support modifying diffs, described later: direct editing and moving changes.
This satisfies Requirement~2.

\MyToolName employs a 3-way diff display based on the intermediate version $s'$ shared by the two consecutive diffs, introduced in \cref{s:motivation}.
A straightforward alternative is to display the two diffs independently, as in \cref{fig:commit1,fig:commit2}.
This removes the need to switch views, but the boundary version $s'$ is then represented twice, and each copy is aligned differently against its own counterpart.
Relating a change across the two diffs therefore still requires re-locating the corresponding line in the other view and joining the two copies mentally, and after each modification, its dual effect, leaving one diff and entering the other, appears in two separate places.
The 3-way diff display eliminates this residual reliance on memory by materializing the correspondence: the single shared copy of $s'$ serves as the common coordinate, each line shows its role in both commits at once, and the dual effect of an edit appears on the same line.
This display also matches the mental model of adjusting boundaries: developers can reason about whether each change belongs before or after the boundary.

\subsection{Comparison}
\MyToolName arranges three versions of source codes side by side in a 3-way diff layout.
The difference between the left and center source code corresponds to the first commit, and the difference between the center and right source code corresponds to the second commit.

\MyToolName colors the background using a combination of line-level and character-level highlighting, consistent with existing diff displays.
The left source code corresponds to the deletion in the first diff, and the right source code corresponds to the addition in the second diff.
Accordingly, as seen in line 295 of \cref{fig:method-overview}, the left and right source code are assigned the same background colors as in a typical diff editor.
In contrast, the center source code corresponds to both the addition in the first diff and the deletion in the second diff.
Lines or characters that represent only an addition or a deletion are displayed with background colors similar to those of a typical diff editor, as seen in line 302 of \cref{fig:method-overview}.
Lines or characters representing both an addition and a deletion are shown with a green and red diagonal stripe pattern, as seen in line 295.
Green denotes an addition and red a deletion, as in a typical diff editor; unmodified content is left uncolored, and the cited line numbers are those shown in the editor gutter in \cref{fig:method-overview}.

\subsection{Modification}\label{s:modification-method}

\MyToolName provides two operations for modifying diffs: direct editing of commit boundaries and moving changes at different granularities.
Direct editing allows boundaries to be placed at any position, but it requires significant effort.
In contrast, moving changes is limited by available granularity, but it requires less effort.
Therefore, moving changes serves as the primary operation, while direct editing is intended for finer modifications such as separating changes within the same line.

\Heading{Direct Editing}
Developers perform direct edits by modifying the center source code.
Because the center source code is the boundary between the two commits, editing it lets developers place the commit boundary at any position.

\Heading{Moving Changes}
Moving changes is available at two levels of granularity: line level and hunk level (groups of consecutive lines).
This operation is inspired by the buttons in IDE diff editors that allow developers to stage changes by hunk.
Line-level movement is also provided because a hunk can span dozens of lines, making hunk-level movement alone too coarse-grained.

\begin{figure}[tb]\centering
  \begin{subfigure}[b]{\linewidth}
    \centering
    \includegraphics[width=.95\linewidth]{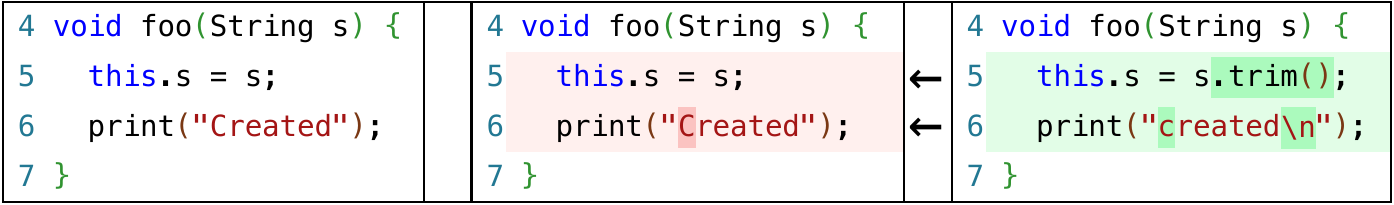}%
    \caption{State before line-level or hunk-level movement.}
    \label{fig:before-move}
  \end{subfigure}

  \vspace{0.5\baselineskip}
  \begin{subfigure}[b]{\linewidth}
    \centering
    \includegraphics[width=.95\linewidth]{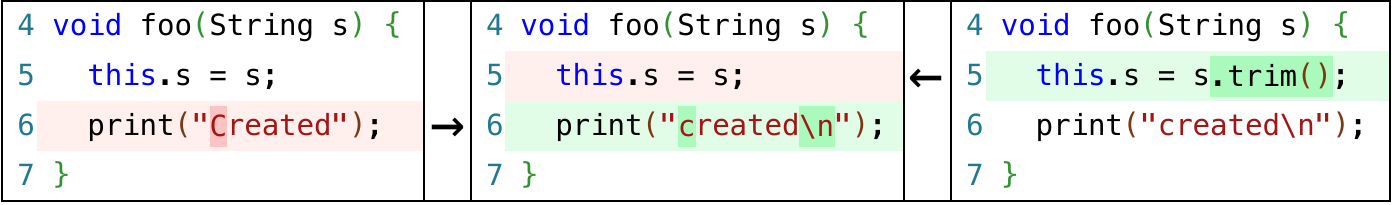}%
    \caption{Moving the change on Line 6 of (a) to the first commit.}
    \label{fig:line-move}
  \end{subfigure}

  \vspace{0.5\baselineskip}
  \begin{subfigure}[b]{\linewidth}
    \centering
    \includegraphics[width=.95\linewidth]{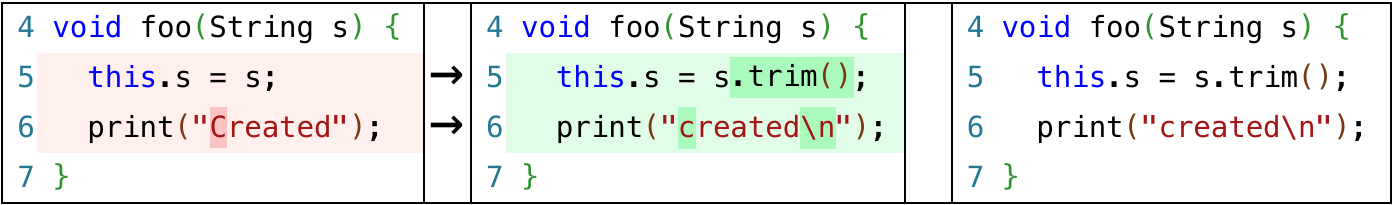}%
    \caption{Moving the hunk on Lines 5--6 of (a) to the first commit.}
    \label{fig:hunk-move}
  \end{subfigure}
  \caption{Modification by line-level or hunk-level movement.}\label{fig:move-example}
\end{figure}

Developers can move changes by clicking the buttons located between the source codes.
These buttons appear for lines or hunks where the content differs when comparing the outer (left or right) source code with the inner (center) source code.
When a button is clicked, the corresponding part of the center source code is replaced with the content from the outer source code.
Because the center version is the boundary, replacing part of it with the left version removes that change from the first diff and adds it to the second, and replacing it with the right version does the reverse.
Moving is thus a structured special case of directly editing the boundary.
Developers can toggle granularity by holding a specific modifier key.

\Cref{fig:move-example} applies the two granularities to the state in \cref{fig:before-move}.
A line-level move sends the change on Line 6 to the first commit, as shown in \cref{fig:line-move}, and a hunk-level move sends the hunk on Lines 5--6 to the first commit, as shown in \cref{fig:hunk-move}.

\subsection{Implementation}\label{s:implementation}

\begin{figure*}[tb]
  \centering
  \includegraphics[width=1\linewidth]{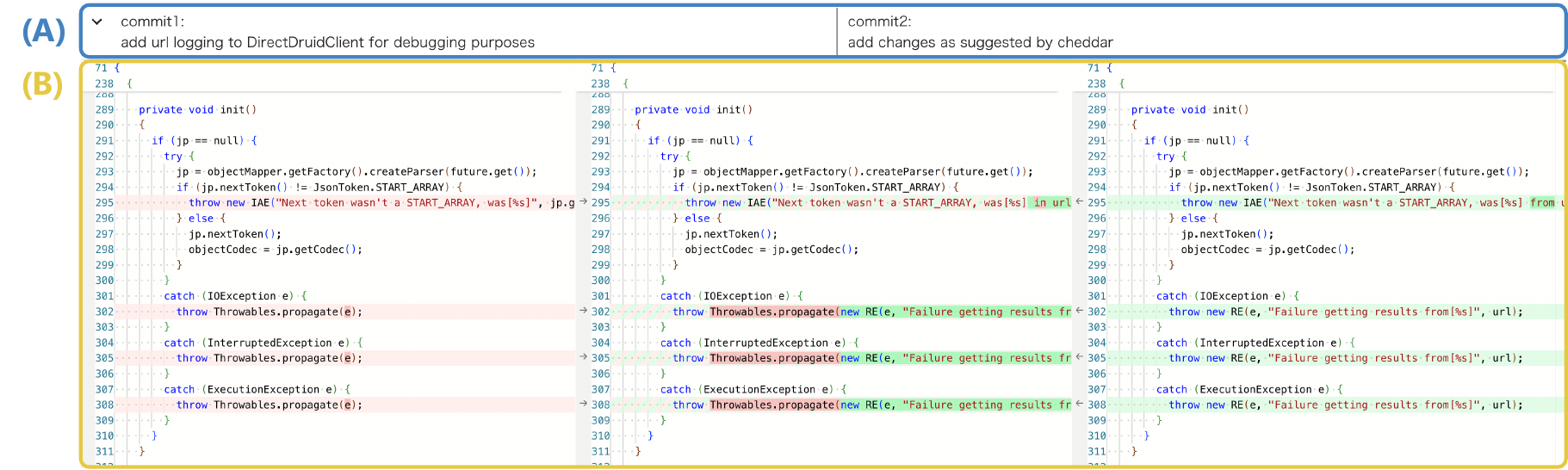}
  \caption{Screenshot of \MyToolName.}
  \label{fig:entire-screen}
\end{figure*}

We implemented \MyToolName as a web application using TypeScript and Nuxt.js.
As shown in \cref{fig:entire-screen}, this tool consists of two parts: (A) a header and (B) a diff adjustment editor.
The header is an area that displays information about the two target commits, including the SHA and message of each commit, as well as a button for toggling the visibility of the messages.
The diff adjustment editor provides the diff adjustment environment based on the 3-way diff display proposed in \cref{s:technique}.

In the following, we describe how \MyToolName constructs the 3-way diff display from its input.
The input consists of three versions of the same file, $s$, $s'$, and $s''$, where $(s, s')$ corresponds to the first commit and $(s', s'')$ to the second; the center version $s'$ is shared by the two commits as their boundary.
The construction proceeds in two sequential stages, one at the line level and one at the character level, whose results are then rendered on the editor.

At the line level, we obtain the three versions of the source code aligned line by line, together with a per-line classification.
First, we compute the line-level diff between $s$ and $s'$ and that between $s'$ and $s''$ using the Myers algorithm \cite{Myers1986} provided by \texttt{jsdiff}\cite{jsdiff}.
Second, using the center version $s'$ as the shared anchor, we align $s$ and $s''$ to it: the first diff maps each line of $s'$ to its corresponding line in $s$, and the second diff maps it to its corresponding line in $s''$.
Third, we equalize the three columns row by row by inserting blank lines: within each changed region the shorter side is padded so that all three versions span the same number of rows, and the padding of the shared center is taken as the larger of the amounts the two diffs require.
This construction follows the same principle as diff3 \cite{diff3}: it anchors on the center version and composes two pairwise diffs rather than introducing a new alignment algorithm.
We reuse this well-understood alignment principle intentionally, concentrating our contribution on the interaction layer.
Finally, we classify each line as \textit{unmodified}, \textit{added}, \textit{removed}, \textit{both}, or \textit{empty}.
A center line is classified as \textit{both} when it is added by the first commit and deleted by the second.
Padding lines are classified as \textit{empty}.

At the character level, we obtain a per-character classification within each hunk.
A hunk is a group of consecutive lines in which either of the two versions is classified as \textit{added}, \textit{removed}, or \textit{both}.
For each hunk, we join its lines into one string per version and compute two character-level diffs with \texttt{diff-match-patch}\cite{diffmatchpatch}: one between $s$ and $s'$ and one between $s'$ and $s''$.
This library applies the Myers algorithm and then a semantic cleanup that improves the readability of the resulting diff.
Each differing character is classified as \textit{added}, \textit{removed}, or \textit{both}, as in the line-level classification.
Because the two character-level diffs are computed independently, they can occasionally disagree on the classification of a shared span; resolving such cases more precisely is left for future work.

Finally, we render the three aligned versions of the source code on three editors placed side by side, implemented with the Monaco Editor\cite{monaco}.
Each editor is decorated with the line-level background colors from the line-level classification and the character-level highlighting from the character-level classification.
The three editors are scroll-synchronized so that corresponding lines remain aligned.
Only the center editor is editable to support direct editing.
Movement buttons are placed on the lines or hunks whose content differs between the inner source code and an outer source code.

\section{Evaluation}\label{s:evaluation}

We set the following Research Questions (RQs) to evaluate the usefulness of the proposed approach \MyToolName from the perspective of a developer adjusting commit boundaries.
\begin{itemize}
  \item \RQ{1} (Accuracy): Does \MyToolName improve the accuracy of adjusting commit boundaries compared to the baseline?
  \item \RQ{2} (Efficiency): Does \MyToolName improve the efficiency of adjusting commit boundaries compared to the baseline?
  \item \RQ{3} (Usability): Does \MyToolName improve developer usability in adjusting commit boundaries compared to the baseline?
\end{itemize}

To address these RQs, we conducted a within-subjects user study.
In this experiment, participants adjusted commit boundaries using both \MyToolName and a baseline tool, and then completed questionnaires.

\subsection{Experimental Setup}

\subsubsection{Tools}

We used VS Code with the GitLens\cite{gitlens} extension installed as a baseline tool.
No purpose-built tool adjusts the boundary between two given consecutive commits, as the tools in \cref{s:related-work} operate on a single composite commit, so the realistic choice for our baseline is a general VCS workflow, which moves changes across an existing boundary through interactive rebasing.
VS Code was ranked as the most widely used IDE in the Stack Overflow Developer Survey 2025\footnote{\UrlSurvey}, so the comparison reflects practical use.
While VS Code provides Git GUI operations as a standard feature, it does not support operations necessary for adjusting commit boundaries, such as interactive rebasing.
GitLens is an extension that complements these missing features with a GUI.

We implemented the approach \MyToolName proposed in \cref{s:technique} and used it in the experiment.
For this implementation, we enabled detailed logging.
The logs include the start and end times of work on specific tasks, as well as modifications made through moving or editing changes.

\subsubsection{Dataset}

We built our own dataset of commit-boundary adjustment tasks, as none is available.
Such a task needs both a misplaced boundary as its starting point and the correct boundary as ground truth for scoring, and cases that offer both are rare in real repositories.
A boundary adjustment is usually made on local commits before pushing, so the misplaced starting state seldom reaches the shared repository, and a misplaced boundary that does remain there rarely carries a recorded correct target to score against.
Finding cases suitable for the experiment in real repositories is therefore challenging, so we instead took clean TLC pairs as ground truth and altered their boundaries to obtain controlled misplaced starting states.
As the source of clean TLC pairs, we used an artificially generated TLC dataset proposed in UTANGO\cite{Li2022}.
CC datasets are either manually verified from real repositories or artificially generated; the manually verified ones\cite{Herzig2013, Herzig2016} are not publicly available, whereas artificially generated ones extract real TLC sequences and recombine them with \texttt{git cherry-pick}.
Of the several available\cite{Partachi2020, Shen2021, Li2022}, only UTANGO met the criteria of the extraction step described later.

\begin{table*}[tb]\centering
  \caption{TLC Pairs Selected as Tasks}\label{t:atomic-commit-pairs}
  \begin{tabular}{cllllcr} \hline
    Group & Task & Repository & Commit Pair & Filename & In-line Separation & Distance\\\hline
    \multirow{5}{*}{A} & A1 & ReactiveX/RxJava & \href{https://github.com/ReactiveX/RxJava/compare/b611684%5E...ffbfb67}{\texttt{b611684}, \texttt{ffbfb67}} & BlockingObservable.java &  & 56\\
    & A2 & square/retrofit & \href{https://github.com/square/retrofit/compare/f78f2fd%5E...5cadd33}{\texttt{f78f2fd}, \texttt{5cadd33}} & RestAdapter.java &  & 127\\
    & A3 & elastic/elasticsearch & \href{https://github.com/elastic/elasticsearch/compare/35d947b%5E...742607e}{\texttt{35d947b}, \texttt{742607e}} & RestIndicesAction.java & $\X$ & 28\\
    & A4 & elastic/elasticsearch & \href{https://github.com/elastic/elasticsearch/compare/83a61ca%5E...c3ccfe0}{\texttt{83a61ca}, \texttt{c3ccfe0}} & BoolFilterParser.java & $\X$ & 3\\\cline{2-7}
    & Total & & & & & 214\\\hline
    \multirow{5}{*}{B} & B1 & elastic/elasticsearch & \href{https://github.com/elastic/elasticsearch/compare/0d2675e%5E...92a6968}{\texttt{0d2675e}, \texttt{92a6968}} & ShardIndexingService.java &  & 215\\
    & B2 & elastic/elasticsearch & \href{https://github.com/elastic/elasticsearch/compare/112935f%5E...b143400}{\texttt{112935f}, \texttt{b143400}} & IpFieldMapper.java &  & 25\\
    & B3 & NationalSecurityAgency/ghidra & \href{https://github.com/NationalSecurityAgency/ghidra/compare/3813583%5E...f15798c}{\texttt{3813583}, \texttt{f15798c}} & VarnodeAST.java & $\X$ & 26\\
    & B4 & square/retrofit & \href{https://github.com/square/retrofit/compare/478e1dd%5E...0f59dc7}{\texttt{478e1dd}, \texttt{0f59dc7}} & SimpleXMLConverter.java & $\X$ & 64\\\cline{2-7}
    & Total & & & & & 330\\\hline
  \end{tabular}
\end{table*}

From this dataset, we extracted tasks usable in the experiment based on the limitations of the proposed approach.
We first limited the \NumCommitSequences TLC sequences in the UTANGO dataset to those of length 2, obtaining \NumCommitPairs commit pairs.
Next, we restricted the pairs to those where the two commits each involve the same single file, obtaining \NumCommitPairsSingleFile commit pairs.

Finally, we performed a visual inspection to select only TLC pairs suitable for the experiment, obtaining \NumProblemsTotal commit pairs.
The first criterion was that the commit message accurately described the changes contained in the TLC.
This is because the commit message is provided to participants as a guideline for the adjusting task in our experiment.
An inaccurate message would prevent participants from identifying the correct state to be achieved.
The second criterion was that the changes were non-trivial, ensuring a meaningful task.

We set task labels for each of the \NumProblemsTotal selected pairs.
We divided the tasks into two groups (A and B) of \NumProblemsPerGroup tasks each to allow each participant to use both tools.
We imposed two constraints on this division.
First, both groups contained the same number of tasks requiring in-line separation (2 out of \NumProblemsIntraLine each).
Second, the per-repository difference in task count between the groups was at most one.
Subject to these constraints, we minimized the inter-group gap in the sum of Levenshtein distances.
Each distance was measured between the original boundary from the TLC pair in the dataset and the commit boundary input to the participant (described later).
\Cref{t:atomic-commit-pairs} shows the selected TLC pairs and the group division.

For the selected TLC pairs, we altered the commit boundaries line by line, the simplest choice for this first step, and used the resulting pairs as input for the participants in the experiment.
First, we mapped the corresponding lines between the two commits and listed the lines where changes occurred in either commit.
From that list, we randomly selected $\BoundaryAlterationRate\%$ of the lines and moved those lines to the opposite commit.
Note that for lines where changes were made in both commits, we moved them together to one of the two commits chosen at random.
Restoring such a line requires character-level separation rather than a line move, which is why some tasks require the in-line separation marked in \cref{t:atomic-commit-pairs}.
These listing and movement operations were repeated until the resulting commit boundaries became syntactically correct.

\subsection{Experimental Process}

We recruited \NumParticipants participants from the authors' affiliated institution, requiring students majoring in computer science with at least two years of Java programming experience.
Each participant received a 3,000 JPY (approx. \$20) Amazon gift card as compensation for their participation.
In accordance with the ethics screening criteria of the authors' affiliated institution, this study was determined not to require a detailed ethical review.
Before starting the task, we provided the participants with preliminary explanations and tutorials through pre-recorded videos, covering the background of the experiment, the usage of the two tools, and the experimental procedure, followed by one practice task with each tool.
To mitigate the difference in required operations between the tools, the video for the baseline tool was given more than twice the length of that for \MyToolName (15 vs.\ 7 minutes).

We applied counterbalancing to the assignment of experimental conditions to minimize order and learning effects.
Each participant used different tools across two sessions and addressed different task groups.
We adopted a $4 \times 4$ Latin square based on Williams' design\cite{williams1949} to eliminate the carryover effect from preceding tasks.
This design ensures that each task appears equally in every position, and that the task preceding any given task varies among participants.

We asked the participants to adjust commit boundaries according to their assignment.
The original TLC commit messages served as the guideline for the adjustment task.
Since the baseline tool does not support detailed logging, we instructed the participants to record their screens and measure the time taken for each task.
The screen recording of one participant was lost when the recording tool crashed.

After the participants finished the adjusting commit boundaries tasks, we asked them to complete a questionnaire.

\subsection{\RQ{1}: Accuracy}

\subsubsection{Study Design}

We use the improvement rate of the center source code as an evaluation metric for accuracy.
Let $s^\mathrm{in}$ be the center source code presented as input to the participant, and $s^\mathrm{out}$ be the center source code obtained by the participant upon completion of the task.
Let $s^\ast$ be the correct center source code, i.e., the original center source code of the TLC pair in the dataset.
Let $d(a,b)$ be the Levenshtein distance between source codes $a$ and $b$.
The distance is calculated in characters, counting whitespace characters such as newlines, tabs, and spaces as ordinary characters, so that differences in formatting are also reflected as errors.
The improvement rate is then defined as follows:
\begin{equation*}
  \mathrm{Imp} = \begin{cases}
    1 - \dfrac{d(s^\mathrm{out}, s^\ast)}{d(s^\mathrm{in}, s^\ast)}
      & \text{if } d(s^\mathrm{out}, s^\ast) \le d(s^\mathrm{in}, s^\ast), \\
    \dfrac{d(s^\mathrm{in},\, s^\ast)}{d(s^\mathrm{out},\, s^\ast)} - 1
      & \text{if } d(s^\mathrm{out}, s^\ast) > d(s^\mathrm{in}, s^\ast).
  \end{cases}
\end{equation*}
$\mathrm{Imp}$ measures how much the Levenshtein distance from the correct answer shrinks relative to the input: $\mathrm{Imp}=1$ indicates a perfect match, $\mathrm{Imp}=0$ no improvement, and a negative value a deterioration.
We treat $\mathrm{Imp}$ as a proxy for the correctness of the adjustment rather than a validated accuracy measure; it scores the character-exact distance to a single reference $s^\ast$, so a functionally correct but differently formatted result still counts as imperfect.
Because both tools are scored by this same metric, it penalizes neither tool over the other and does not bias their comparison.

We evaluate the difference between the tools based on the improvement rate.
We treat the participant as the primary unit of analysis: for each participant, we take the median of the improvement rates over the \NumProblemsPerGroup tasks performed with each tool, obtaining one pair of values per participant, and perform a Wilcoxon signed-rank test on the resulting \NumParticipants within-participant pairs.
Because of the counterbalanced design, the two values in a pair come from different task groups; counterbalancing (a Williams square) balances task difficulty across participants rather than within each pair.
As a complementary robustness view, we also report a per-task analysis that pairs the median improvement rate of each tool on each of the \NumProblemsTotal tasks.
If significant, we evaluate the effect size using Cliff's delta\cite{Romano2006}.

\subsubsection{Results and Discussion}

\Cref{fig:improvement-distribution} shows the distribution of the improvement rate for each tool.
Blue and red boxes represent the baseline tool and \MyToolName, respectively.
The left plot shows the overall distribution: the median improvement rate is higher for \MyToolName than for the baseline tool.
The right plot shows the per-task distribution, where no clear difference between the tools is observed.

This higher median for \MyToolName, however, is not statistically significant.
The participant-level Wilcoxon signed-rank test yielded a p-value of \PValueAccuracy, failing to show a significant difference at the 5\% significance level; the complementary per-task analysis agrees ($p = \PValueAccuracyPerTask$).

\Cref{fig:improvement-distribution} shows that the improvement rate is low for tasks requiring in-line separation.
Specifically, the improvement rate is low for tasks A3, B3, and B4, all of which require this type of separation.
We examined the center source code for these tasks.
Direct editing makes in-line separation possible, as described in \cref{s:modification-method}; some participants used it to separate the changes correctly, while others hardly used it or applied it without achieving a correct separation.
The capability alone may therefore not be enough; the difficulty lies less in editing than in noticing that a single line mixes changes from both commits, so support that makes such lines easier to spot could be useful.
In addition, for task A3, which had the lowest improvement rate, the participants failed to notice the need for separation; recognizing it required domain knowledge of the code being changed, such as a non-obvious default the Joda-Time library applies for an omitted constructor argument.

\Conclusion{%
  No statistically significant difference was found in the accuracy of the two tools.
  Additionally, across all tasks, the improvement rate tended to be lower for tasks requiring in-line separation compared to those that did not.
}

\subsection{\RQ{2}: Efficiency}

\subsubsection{Study Design}

We use the time required to adjust commit boundaries as an evaluation metric for efficiency.
\MyToolName is timed in the logs from the start button, which immediately displays the three-pane diff, to the end button pressed upon completing the adjustment.
The baseline is timed by one of the authors from the first diff display to rebase or stash completion on the screen recording, or from self-report for the participant whose recording was lost.
Both intervals thus run from the first display of the diff to the completion of the adjustment in each tool.
They are not perfectly symmetric: the baseline includes executing the rebase or stash, whereas \MyToolName does not produce actual commits; we believe this discrepancy is small relative to the observed difference in adjustment time.
We report no inter-rater reliability; the baseline interval excludes pre-display reading and post-task verification, so imprecision in its endpoints underestimates rather than inflates the baseline times.

In addition, we counted the number of screen switches with the baseline tool from the same screen recordings to investigate the cause of the difference in adjustment time.
We defined a switch as a transition of the view displayed in the editor area, e.g., between diff editors and the text editor.
This analysis covers the \NumParticipantsRecorded participants with an intact recording.
We also draw on the participants' free-text questionnaire responses for qualitative insight into the difference in adjustment time.

We evaluate the difference between tools in terms of the adjustment time.
As in \RQ{1}, we treat the participant as the primary unit: for each participant, we take the median of the adjustment times over the \NumProblemsPerGroup tasks performed with each tool and apply the Wilcoxon signed-rank test and Cliff's delta\cite{Romano2006} to the \NumParticipants within-participant pairs.
As a complementary view, we also report the per-task analysis that pairs the median adjustment time of each tool on each task.

\subsubsection{Results and Discussion}

\begin{figure}[tb]\centering
  \includegraphics[width=1\linewidth]{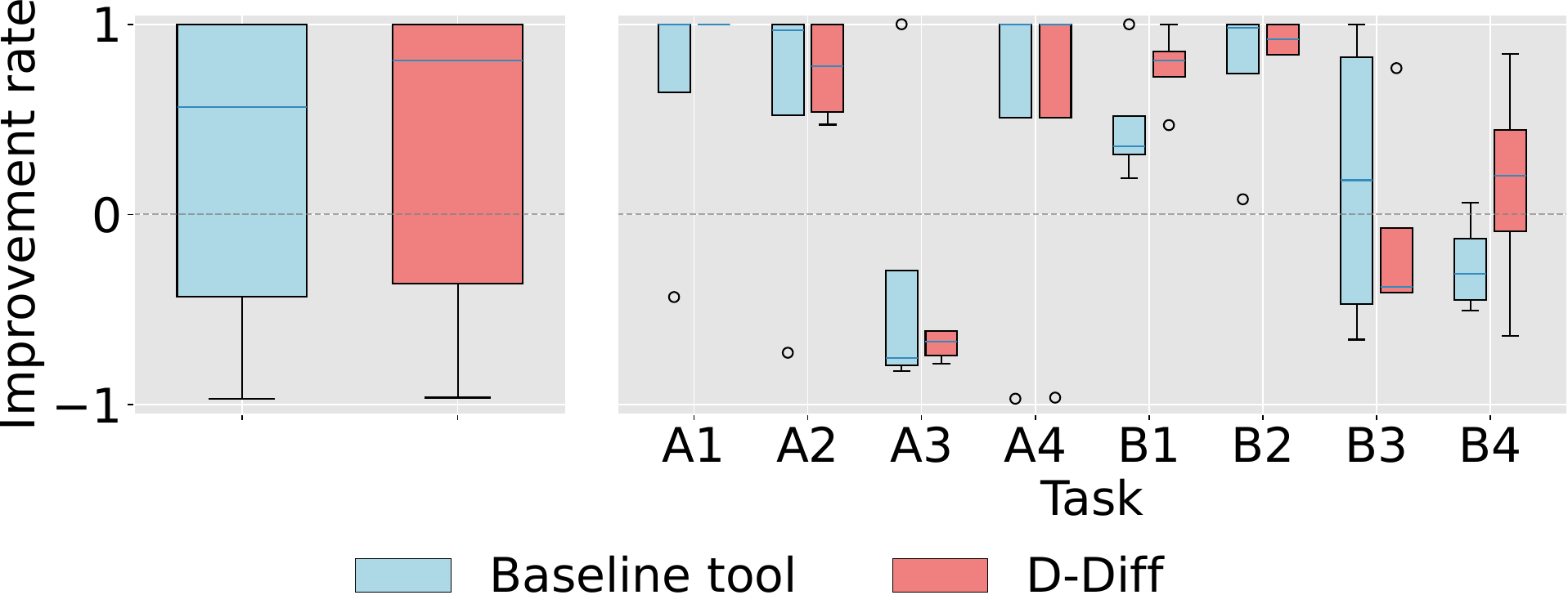}
  \caption{Improvement rate for each tool (Left: overall, Right: per task).}\label{fig:improvement-distribution}
\end{figure}

\begin{figure}[tb]\centering
  \includegraphics[width=1\linewidth]{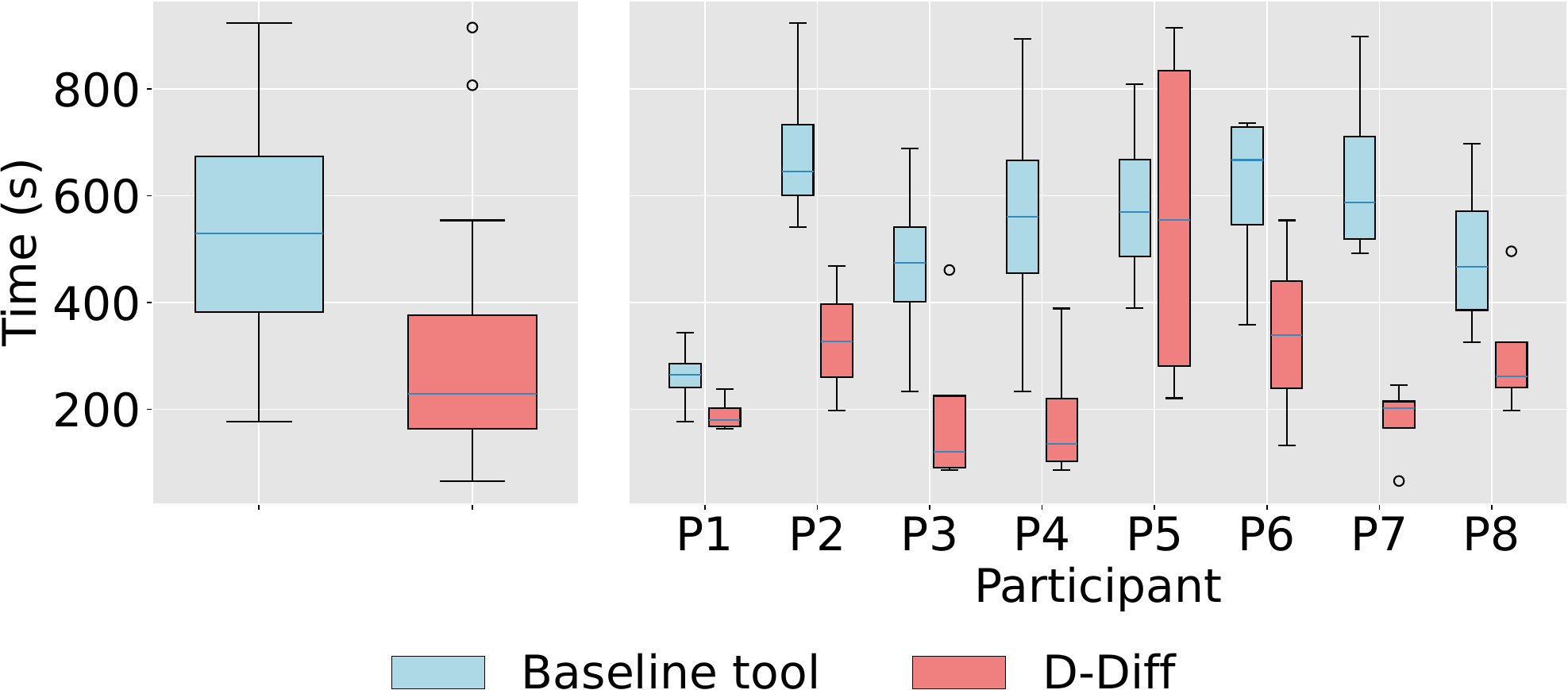}
  \caption{Time taken by tool (Left: overall, Right: per participant).}\label{fig:task-time-distribution}
\end{figure}

\Cref{fig:task-time-distribution} shows the distribution of adjustment time for each tool.
Blue and red boxes represent the baseline tool and \MyToolName, respectively.
The left plot shows the overall distribution of adjustment time.
The median adjustment time is shorter for \MyToolName than for the baseline tool.
The median times for the baseline tool and \MyToolName were \MedianTimeBaseline seconds and \MedianTimeDdiff seconds, respectively, with \MyToolName reducing the time by \TimeReductionSec seconds (\TimeReductionPct\%).
The right plot groups the adjustment time by participant, in contrast to the per-task grouping used for accuracy in \cref{fig:improvement-distribution}, because the efficiency advantage is consistent across participants.
Every participant was faster overall with \MyToolName than with the baseline tool.

\MyToolName significantly reduces the adjustment time compared to the baseline.
The participant-level Wilcoxon signed-rank test yielded a p-value of \PValueEfficiency, indicating a statistically significant difference at the 5\% significance level.
The complementary per-task analysis agrees ($p = \PValueEfficiencyPerTask$); although the per-task median for task B4 favors the baseline tool, this reversal appears only in the per-task view, since every participant who used \MyToolName on B4 was still faster overall.
In addition, Cliff's delta is \CliffsDelta, indicating a \textit{\CliffsDeltaInterpret} effect size.

\begin{figure}[tb]
  \centering
  \includegraphics[width=0.8\linewidth]{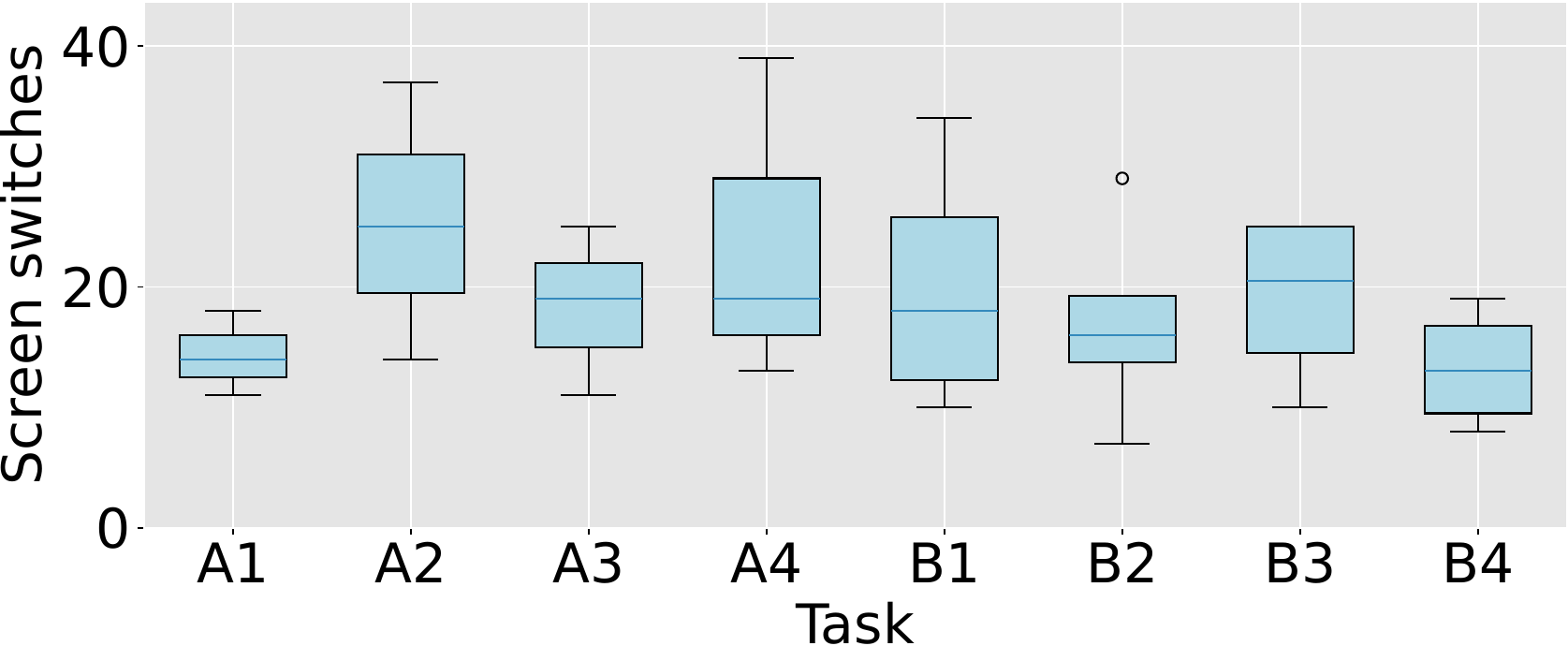}
  \caption{Number of screen switches per task with the baseline tool.}
  \label{fig:screen-switches}
\end{figure}

The questionnaire responses suggest that part of the difference in adjustment time is attributable to a difference in cognitive load.
\MyToolName may have reduced cognitive load by consolidating two diffs into a single screen.
\Cref{fig:screen-switches} shows the distribution of the number of screen switches with the baseline tool for each task.
The participants switched screens \MeanScreenSwitches times per task on average, whereas \MyToolName has a single screen by design, so switching does not arise.
The median number of screen switches ranged from \MinProblemMedianSwitches to \MaxProblemMedianSwitches across the \NumProblemsTotal tasks, indicating that frequent screen switching was required in every task rather than only in a few difficult ones.
We treat these switch counts as descriptive corroboration of the cognitive-load hypothesis rather than an independent test, since they come from the same recordings as the baseline times and from the only condition in which switching occurs.
One participant noted that \textit{``\MyToolName had a simple UI that consolidated necessary information into one screen, making it easy to operate,''} while another said the baseline tool \textit{``required operations across multiple screens, which really trained my working memory [required effort to remember].''}
These responses suggest that the baseline tool imposed a high memory load from referencing multiple screens, which \MyToolName mitigates.

The two tools also appear to differ in the intuitiveness of their operations: one participant reported that \textit{``In the baseline tool, there were several points where I didn't know how to operate without looking at the slides, which took time,''} whereas another found that \textit{``With \MyToolName, I was able to grasp the task to be done and perform corrections intuitively, even for the first time.''}
These responses suggest that while the baseline tool required trial and error before reaching the desired operation, \MyToolName made it easy to identify the next step.
These differences in cognitive load and operability likely contributed to the shorter adjustment time with \MyToolName.
At the same time, because the boundary alteration moves whole lines, the injected errors are largely what \MyToolName's move operations address, so part of this efficiency advantage may reflect the task design rather than the tool in general.

\Conclusion{%
  The adjustment time was significantly shorter with \MyToolName than with the baseline tool ($p = \PValueEfficiency$, \CliffsDeltaInterpret).
  Furthermore, questionnaire responses suggest that the reduced cognitive load and improved operability of \MyToolName contributed to the reduction in adjustment time.
}

\subsection{\RQ{3}: Usability}

\subsubsection{Study Design}

Usability is evaluated based on Likert-scale questionnaire items covering three aspects.
The first aspect assesses the ease of use of each tool with five items: comparing the changes included in the commits, keeping track of the state of the source code being modified, modifying by moving, modifying by editing, and overall ease of use.
The second aspect assesses the ease of use of five individual features of \MyToolName: the parallel display of the two diffs, the background coloring of the center source code, line-level movement, hunk-level movement, and direct editing.
The third aspect directly compares the two tools in terms of perceived task completion time, mental burden, and willingness to use.
Free-text fields asked for the reasons behind these ratings.
These items are author-designed, and \RQ{3} reports the response distributions rather than a confirmatory test, surfacing perceived usability.

\subsubsection{Results and Discussion}

\begin{figure}[tb]\centering
  \begin{subfigure}[b]{\linewidth}\centering
    \includegraphics[width=\linewidth]{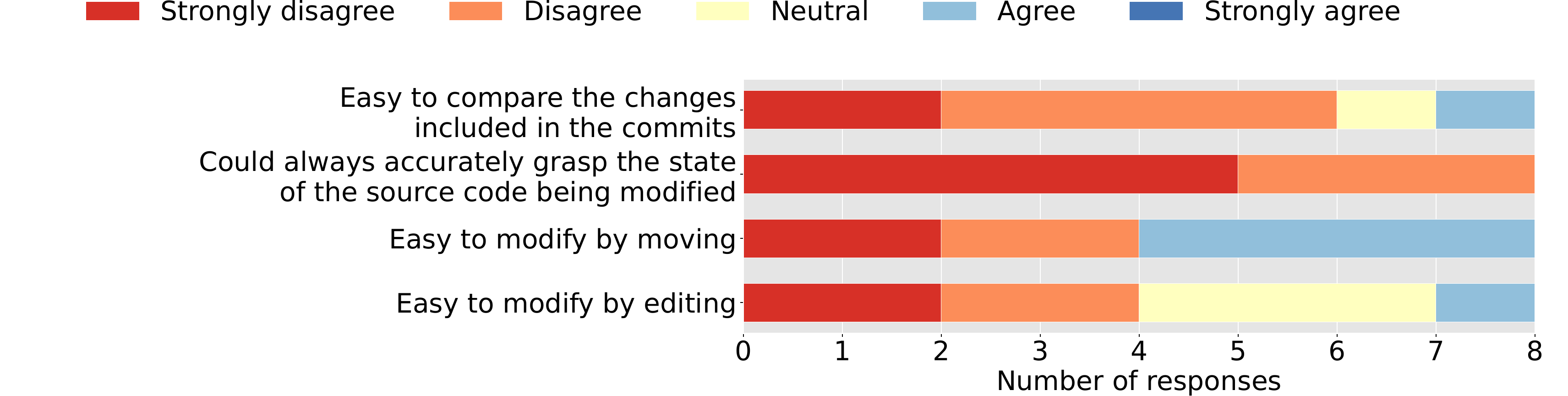}%
    \caption{Usability of the baseline tool.}\label{fig:tool_p_usability}
  \end{subfigure}\vspace{1em}

  \begin{subfigure}[b]{\linewidth}\centering
    \includegraphics[width=\linewidth]{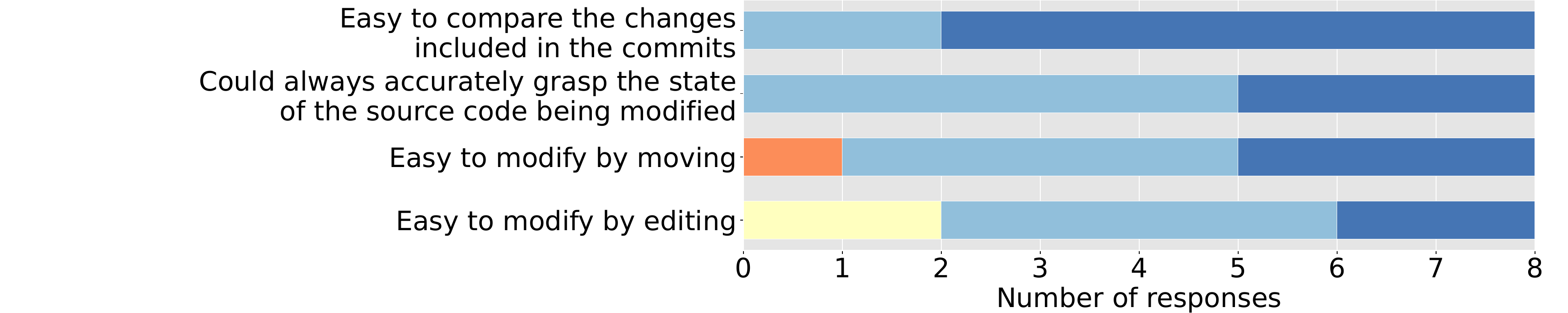}%
    \caption{Usability of \MyToolName.}\label{fig:tool_q_usability}
  \end{subfigure}\vspace{1em}

  \begin{subfigure}[b]{\linewidth}\centering
    \includegraphics[width=\linewidth]{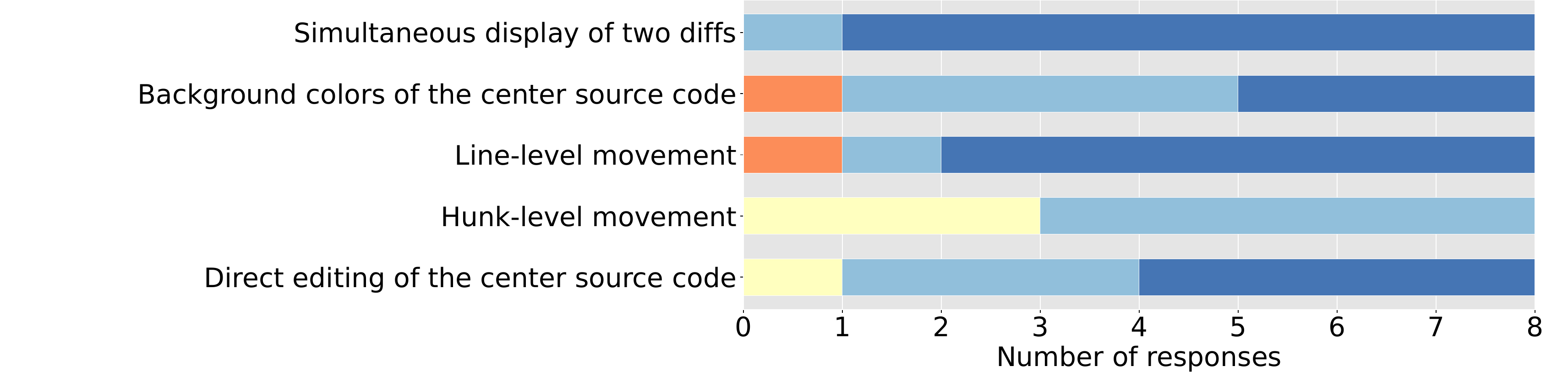}%
    \caption{Individual features of \MyToolName.}\label{fig:tool_q_features}
  \end{subfigure}\vspace{1em}

  \begin{subfigure}[b]{\linewidth}\centering
    \includegraphics[width=\linewidth]{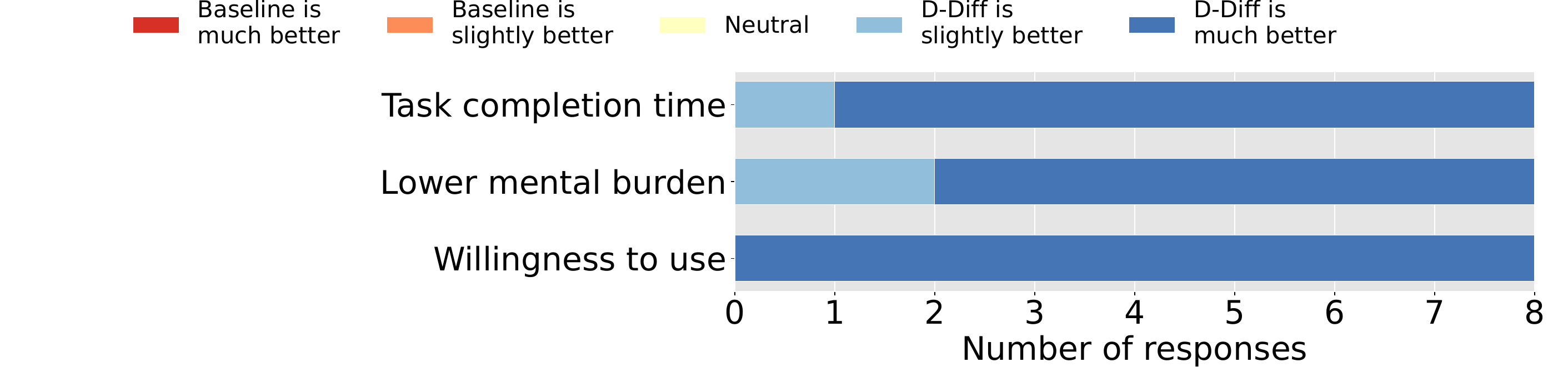}%
    \caption{Comparison between tools.}\label{fig:tool_comparison}
  \end{subfigure}
  \caption{Questionnaire results regarding usability.}\label{fig:usability-survey}
\end{figure}

\Cref{fig:tool_p_usability,fig:tool_q_usability} show the questionnaire results regarding usability for the baseline tool and \MyToolName, respectively.
The vertical axis represents each question regarding usability, and the horizontal axis represents the number of Likert scale responses.
\MyToolName has a higher percentage of positive responses for all questions.
\MyToolName is therefore easier to use than the baseline tool in terms of comparing changes, understanding the source code being modified, and modifying through moving or editing.

\Cref{fig:tool_q_features} shows the questionnaire results regarding the ease of use of each function of \MyToolName.
The vertical axis represents each function.
The percentage of positive responses exceeds half for all features.
All participants ``agreed'' on the simultaneous display of two diffs and line-level movement, while positive responses for hunk-level movement were comparatively fewer.
Each feature of \MyToolName is therefore generally considered easy to use.
The free-text responses indicate that the small scale of the experimental tasks was a reason why hunk-level movement received fewer positive responses, as one participant explained: \textit{``I didn't use it because the diffs in the experiment were small-scale.''}
They also requested clearer background colors for deletions and additions.

\Cref{fig:tool_comparison} shows the questionnaire results regarding the comparison of ease of use between tools.
The vertical axis represents each question.
\MyToolName received a high percentage of positive responses across all questions.
Participants felt that \MyToolName reduced both the perceived task completion time and the mental burden in adjusting commit boundaries, and expressed willingness to use it.

\Conclusion{%
  \MyToolName was consistently rated higher in usability than the baseline tool.
  Participants found it easier to compare and modify changes, and would use it in practice.
}

\subsection{Threats to Validity}

\subsubsection{Internal Validity}
The effect of the 3-way diff display alone may not have been isolated due to the functional differences between \MyToolName and the baseline tool.
For example, the baseline tool supports resetting commits and resolving conflicts, which \MyToolName does not provide, and the two tools require different operations to achieve an adjustment.
The observed differences may therefore reflect the integrated environment as a whole rather than the 3-way display alone; \RQ{2} thus evaluates that environment against a realistic workflow, not the display in isolation.
Comparing \MyToolName with a variant that shows the two diffs in separate views would isolate the effect of the display and is left for future work.
In addition, the participants may have guessed which tool was proposed, partly because the two tutorials differed in length, and answered the questionnaire in its favor; this demand characteristic most affects \RQ{3} and least affects the \RQ{2} timing.
Moreover, most participants self-reported low proficiency in interactive rebase, so their limited familiarity with the baseline workflow may partly explain the large time difference observed in \RQ{2}, although the tutorial and practice task for the baseline tool were designed to mitigate this gap.

\subsubsection{Construct Validity}
The improvement rate based on character-level Levenshtein distance may not fully capture the actual accuracy of the adjustment results, as it cannot capture semantic differences.
In addition, because the injected errors are line-level, the tasks favor move-based correction, so the efficiency advantage may not extend to adjustments that moving alone cannot perform.

\subsubsection{Conclusion Validity}
The primary statistical tests pair the per-participant median of each tool over the \NumProblemsPerGroup tasks, yielding \NumParticipants within-participant pairs, and we report the per-task analysis only as a complementary view.
Because each participant uses each tool on a different task group, the two members of a pair are not matched on the same tasks; counterbalancing balances task difficulty across participants rather than within each pair, and this confound cannot be removed analytically.
A crossed mixed-effects model that separates participant and task effects would be preferable, but its variance components cannot be reliably estimated with only \NumParticipants participants and \NumProblemsTotal tasks.
The power to detect small effects is therefore limited, and the absence of a significant difference in accuracy does not imply that the two tools are equivalent.

\subsubsection{External Validity}
The number of participants is limited to \NumParticipants, which may not be representative of all developers.
While students majoring in computer science served as participants, prior studies have shown that the performance of students and professional developers is comparable\cite{Salman2015, Falessi2018}.
In addition, further verification is needed to generalize the findings to other languages, since only four Java repositories were used.
Moreover, the tasks were created by randomly altering the boundaries of artificially generated TLC pairs; real tangled commits may require more semantic judgment, and the findings may not directly generalize to them.
Finally, we argue for the relevance of this adjustment scenario but do not empirically measure how often it arises in practice.

\section{Conclusion}\label{s:conclusion}

We proposed \MyToolName, an interactive diff adjustment environment that integrates the comparison and modification of diffs.
\MyToolName enables developers to compare the diffs without relying on their memory by visualizing two diffs simultaneously based on a 3-way diff display.

We evaluated its usefulness through a user study with \NumParticipants participants.
The results showed that \MyToolName was significantly superior to the baseline tool in terms of efficiency.
The median adjustment time was \MedianTimeDdiff seconds for \MyToolName compared to \MedianTimeBaseline seconds for the baseline tool, representing a reduction of approximately \TimeReductionPct\%.
\MyToolName also received a higher percentage of positive responses across all usability questionnaire items, with participants expressing willingness to use it in practice.
No statistically significant difference in accuracy was found between the tools ($p = \PValueAccuracy$).

Future work on the tool itself includes comparing related changes across different files, extending to longer commit sequences, integrating \MyToolName into IDEs, and supporting fine-grained modifications such as in-line separation.
On the evaluation side, the user study could be extended to practitioners.
Beyond commit boundary adjustment, the core interaction of directly editing a shared boundary state could also support splitting a single commit, redistributing changes between any two versions, and curating the outputs of automatic untangling.

The tool and evaluation data are available \cite{dataset}.

\section*{Acknowledgments}
This work was partly supported by JSPS KAKENHI (JP26K02889, JP23K24823, JP26K02888, JP25K03102, JP25H01125, and JP24H00692).

\Heading{Declaration of AI Usage}
During the preparation of this work, the authors used Claude and Nani to improve presentation, including illustration, text readability, and language; they reviewed, edited, and take full responsibility for all content.

\bibliographystyle{IEEEtran}
\bibliography{references}

\end{document}